\documentclass[12pt,oneside,letterpaper]{article}
\usepackage{epsfig}
\usepackage{graphicx}

\def \munit {{$ GeV/c^2$}}               

\begin{document}

\title{An Experimental Overview of Gluonic Mesons}
\author{Curtis A. Meyer \\
Carnegie Mellon University, \\  Pittsburgh, PA 15213}
\maketitle
\begin{abstract}
In this paper, I review the experimental situation for both 
glueballs and hybrid mesons. Theoretical expectations are discussed,
and a survey of what is known about hybrid mesons and glueballs is
undertaken. Good experimental evidence exists for both states with 
exotic quantum numbers and a glueball which is mixed with the nearby
mesons, but a full understanding of these still requires additional
information.
\end{abstract}
\section{Introduction}
\hspace{5mm}
Gluonic mesons are in the broadest sense a $q\bar{q}$ system in which
the gluonic field contributes directly to the quantum numbers of the 
meson. In terms of the simple quark model, all quantum numbers of mesons
are determined by the $q\bar{q}$ alone. However, Quantum Chromo Dynamics,
(QCD) indicates that this picture is not complete. Lattice QCD calculations
predict that both purely gluonic states, (glueballs), and states with the
gluonic field carrying angular momentum, (hybrids) should exist. Beyond
the lattice, most models which explain observed phenomena also predict
such gluonic mesons to be present. A number of the hybrid states are 
predicted to have $J^{PC}$ quantum numbers which are not accessible
to simple $q\bar{q}$ systems, the so-called \emph{exotic}s.

This article will review the experimental situation for gluonic excitations.
Of particular interest are states with exotic quantum numbers where two 
candidates exist. Hybrids with normal quantum numbers are more difficult
to discern, as they are likely to mix with nearby normal mesons. It is
only through detailed studies of decay and production that they can 
be identified. Finally, evidence exists for a $J^{PC}=0^{++}$ glueball that
is strongly mixed with the nearby scalar mesons.  

\section{The Spectrum of $q\bar{q}$ Mesons}
Within the picture of the quark model, mesons are $q\bar{q}$ pairs which have
been combined with  spin, \textbf{S}, orbital angular momentum,
\textbf{L}, and a possible radial excitation. \textbf{S} can be either 
$0$ or $1$, while \textbf{L} can be any non-negative integer. The quantum 
numbers of the allowed states which are conserved by the strong interaction  
can be built up from these as given as follows:
\begin{eqnarray*}
\mathrm{Total\,\,Spin:\,\,\,} & \mathbf{J}\,=\, & \mathbf{L}\oplus\mathbf{S}
= \mid \mathbf{L}-\mathbf{S}\mid \cdots \mid \mathbf{L}+\mathbf{S}\mid \\
\mathrm{Parity}:         & \mathbf{P}\,=\, & (-1)^{L+1} \\
\mathrm{C-Parity}:       & \mathbf{C}\,=\, & (-1)^{L+S} \\
\mathrm{G-Parity}:       & \mathbf{G}\,=\, & (-1)^{L+S+I} 
\end{eqnarray*}
The light-quark mesons are built up from $u$, $d$, $s$ and their 
antiquarks. This yields nine possible $q\bar{q}$ combinations 
for each set of quantum numbers, (nonets). For the allowed 
values of \textbf{L} and \textbf{S}, Table~\ref{tab:mesons} lists the 
nonets of mesons that can be formed. The first listed is the $I=1$ member,
the second and third are the two $I=0$ members, and the last is the 
$I=\frac{1}{2}$ member which contains non-zero strangeness. The mass in 
the last column is the approximate experimental mass of the lightest 
$I=0$ member of the nonet. If one looks at the $J^{PC}$ quantum
numbers listed in Table~\ref{tab:mesons}, the following values are
missing: $0^{--},0^{+-},1^{-+},2^{+-},3^{-+},\ldots$. Anything identified 
with one of these quantum numbers falls outside of the normal $q\bar{q}$ 
picture of the quark model.
\begin{table}[h!]\centering
\begin{tabular}{ccclc} \hline
\textbf{L} & \textbf{S} & $\mathbf{J}^{PC}$ & Particles & Mass \\ \hline
$0$ & $0$ & $0^{-+}$ & $\pi$,$\eta$,$\eta\prime$,$K$          &$0.5$\munit{}\\
    & $1$ & $1^{--}$ & $\rho$,$\omega$,$\phi$,$K^{*}$         &$0.8$\munit{}\\
$1$ & $0$ & $1^{+-}$ & $b_{1}$,$h_{1}$,$h_{1}\prime$,$K_{1}$  &$1.2$\munit{}\\
    & $1$ & $0^{++}$ & $a_{0}$,$f_{0}$,$f_{0}$,$K_{0}$        &$1.4$\munit{}\\
    & $1$ & $1^{++}$ & $a_{1}$,$f_{1}$,$f_{1}$,$K_{1}$        &$1.3$\munit{}\\
    & $1$ & $2^{++}$ & $a_{2}$,$f_{2}$,$f_{2}\prime$,$K_{2}$  &$1.3$\munit{}\\
$2$ & $0$ & $2^{-+}$ & $\pi_{2}$,$\eta_{2}$,$\eta_{2}$,$K_{2}$&$1.7$\munit{}\\
    & $1$ & $1^{--}$ & $\rho$,$\omega$,$\phi$,$K^{*}$         &$1.7$\munit{}\\
    & $1$ & $2^{--}$ & $\rho_{2}$,$\omega_{2}$,$\phi_{2}$,$K^{*}$ & \\
    & $1$ & $3^{--}$ & $\rho_{3}$,$\omega_{3}$,$\phi_{3}$,$K^{*}$ &$1.7$
\munit{}\\
$3$ & $0$ & $3^{+-}$ & $b_{3}$,$h_{3}$,$h_{3}$,$K_{3}$ & \\
    & $1$ & $2^{++}$ & $a_{2}$,$f_{2}$,$f_{2}$,$K_{2}$ & \\
    & $1$ & $3^{++}$ & $a_{3}$,$f_{3}$,$f_{3}$,$K_{3}$ & \\
    & $1$ & $4^{++}$ & $a_{4}$,$f_{4}$,$f_{4}\prime$,$K_{4}$ & $2.0$\munit{}\\
$4$ & $0$ & $4^{-+}$ & $\pi_{4}$,$\eta_{4}$,$\eta_{4}$,$K_{4}$& \\
    & $1$ & $3^{--}$ & $\rho_{3}$,$\omega_{3}$,$\phi_{3}$,$K_{3}$ &$2.25$
\munit{}\\
    & $1$ & $4^{--}$ & $\rho_{4}$,$\omega_{4}$,$\phi_{4}$,$K_{4}$ & \\
    & $1$ & $5^{--}$ & $\rho_{5}$,$\omega_{5}$,$\phi_{5}$,$K_{5}$ &$2.35$
\munit{}\\ 
\hline
\end{tabular}
\caption{\label{tab:mesons}
The quantum numbers of the nonets built up from the allowed \textbf{L} 
and \textbf{S} quantum numbers. The quoted mass is that for the lightest
isoscalar state in the nonet.}
\end{table}

\section{The Spectrum of Gluonic Mesons}

Hybrid meson quantum numbers can be predicted within the \emph{flux-tube
model}~\cite{isgur-85}. In this picture, the gluonic field forms a flux-tube 
between the $q\bar{q}$ pair. In it ground state, the tube carries no angular 
momentum, but it can be excited. The lowest excitation is $L=1$ rotation which
contains two degenerate states, (clock-wise and counter-clock-wise rotations).
Linear combinations of these can be taken such that the tube behaves as
if it has $J^{PC}=1^{-+}$ or $J^{PC}=1^{-+}$. Adding these to the
$L=0$ mesons, the quantum numbers listed in Table~\ref{tab:hybrid} are
obtained. What is of particular interest is that three of the $J^{PC}$s,
$0^{+-}$, $1^{-+}$ and $2^{+-}$ correspond to non $\bar{q}q$ combinations.
\begin{table}[h!]\centering
\begin{tabular}{|l|l|ccc|} \hline
$L=0$,$S=0$ \,\,\, $J^{PC}=$ & 
$1^{++}$, $1^{--}$ & $a_{1}$,$f_{1}$ & $\rho$, $\omega$
&   \\
$L=0$,$S=1$ \,\,\, $J^{PC}=$& 
$0^{-+}$,$1^{-+}$,$2^{-+}$  & 
$\pi$,$\eta$ &  $\pi_{1}$,$\eta_{1}$ & $\pi_{2}$,$\eta_{2}$ \\
            & $0^{+-}$, $1^{+-}$, $2^{+-}$ &
$b_{0}$,$h_{0}$ & $b_{1}$,$h_{1}$ & $b_{2}$, $h_{2}$ \\ \hline
\end{tabular}
\caption{\label{tab:hybrid}
The $J^{PC}$ quantum numbers of hybrid mesons in the flux-tube 
picture. The given $L$ and $S$ couple with the flux tube to produce the 
listed quantum numbers. The last columns lists the names of the particles 
in the nonets, the $\pi_{1}$, $b_{0}$ and $b_{2}$ correspond to exotic
quantum-number nonets.}
\end{table}

\begin{figure}[h!]\centering
\includegraphics[width=0.9\textwidth]{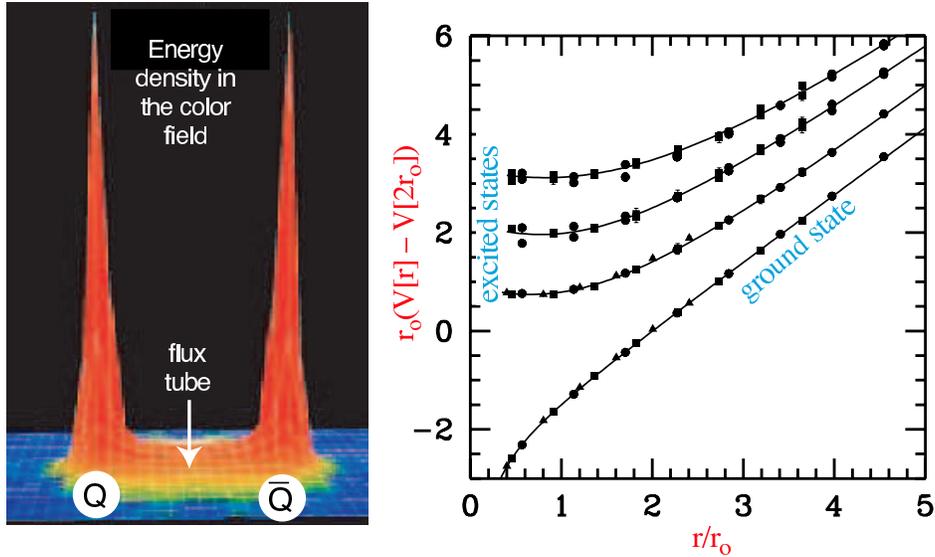}
\caption[]{\label{fig:hybrid}
Lattice calculation of the hybrid potential~\cite{morningstar-99}.}
\end{figure}

Within the flux-tube model, all eight hybrid nonets are degenerate.
The  Lattice also predicts the existence of a flux-tube forming between a
heavy quark-antiquark pair as seen in Figure~\ref{fig:hybrid} left. It is
also possible to calculate the potentials for the ground state and excited 
states of the flux-tube~\cite{morningstar-99} as seen in 
Figure~\ref{fig:hybrid} right. Lattice predictions for hybrid mesons masses
are shown in Table~\ref{tab:lattice-hybrid}. The exotic $1^{-+}$ nonet 
is the lightest state with a mass in the range of $1.8$ to $2$\munit . 
The splitting between the the $1^{-+}$ and $0^{+-}$ nonets is predicted to 
be about $0.2$\munit , (with large errors~\cite{morningstar-01}). 
\begin{table}[h!]\centering
\begin{tabular}{|ll|ll|}\hline
\multicolumn{2}{|c|}{Light Quark $1^{-+}$} &
\multicolumn{2}{c|}{Charmonium $1^{-+}$} \\
Reference & Mass \munit{} & Reference & $\Delta M$ \munit{} \\ \hline
UKQCD~\cite{lacock-97} & $1.87\pm 0.20$       & 
MILC~\cite{bernard-97} & $1.34\pm 0.08\pm 0.20$ \\
MILC~\cite{bernard-97} & $1.97\pm 0.09\pm 30$ & 
MILC~\cite{bernard-99} & $1.22\pm 0.15$ \\
MILC~\cite{bernard-99}  & $2.11\pm 0.13$       & 
\cite{manke-99} & $1.323\pm0.130$ \\
LaSch~\cite{lacock-99} & $1.9\pm 0.20$        & 
\cite{juge-99} & $1.19$ \\
\cite{zhong-02} & $2.013\pm 0.026\pm 0.071$ & & \\ \hline
\end{tabular}
\caption{\label{tab:lattice-hybrid}
Recent results for  $1^{-+}$ hybrid meson masses. }
\end{table}

Predictions for the widths and decays of hybrids are based on model 
calculations with the results of recent work~\cite{page-99} given in 
Tables~\ref{tab:exotic-hybrid-widths} and~\ref{tab:non-exotic-hybrid-widths}.
As can be seen, the width predictions are fairly open. Most of the $0^{+-}$ 
exotic nonet are expected to be quite broad. However, both the $2^{+-}$ and 
the $1^{-+}$ nonets are expected to be much narrower. The non-exotic hybrids
will be more difficult to disentangle as they are likely to mix with nearby 
normal $q\bar{q}$ States. The expected decay modes of hybrids involve 
daughters that in turn decay. This makes the overall reconstruction more 
complicated , with final states involving from four to seven pseudoscalar 
mesons.

However,  these decays can be used as a guideline when looking for the 
states. Almost all models of hybrid mesons predict that the ground state 
ones will not decay to identical pairs of S-wave mesons, and that the decays 
to an $(L=0)(L=1)$ pair is favored. The one unit of angular momentum in the 
flux--tube remains in the internal orbital angular momentum of one of the
daughter  $q\bar{q}$ pairs. 
\begin{table}[h!]
\begin{tabular}{|cc|cc|c|}\hline
Particle & $\mathbf{J^{PC}}$ & \multicolumn{2}{c|}{Total Width $MeV$} & 
Large Decays\\ 
         &                   & \cite{page-99} & \cite{isgur-85a} & \\ \hline
$\pi_{1}$   & $1^{-+}$ &  $81-168$ & $117$ & 
$b_{1}\pi$, $\rho\pi$, $\eta(1295)\pi$\\
$\eta_{1}$  & $1^{-+}$ &  $59-158$ & $107$ &
$a_{1}\pi$, $\pi(1300)\pi$ \\
$\eta'_{1}$ & $1^{-+}$ &  $95-216$ & $172$ &
$K_{1}(1400)K$, $K_{1}(1270)K$, $K^{*}K$ \\ \hline
$b_{0}$     & $0^{+-}$ & $247-429$ & $665$ &
$\pi(1300)\pi$, $h_{1}\pi$ \\
$h_{0}$     & $0^{+-}$ & $59-262$  & $94$  &
$b_{1}\pi$ \\
$h'_{0}$    & $0^{+-}$ & $259-490$ & $426$ &
$K(1460)K$, $K_{1}(1270)K$ \\ \hline
$b_{2}$     & $2^{+-}$ &    $5-11$ & $248$ &
$a_{2}\pi$, $a_{1}\pi$, $h_{1}\pi$ \\
$h_{2}$     & $2^{+-}$ &    $4-12$ & $166$ &
$b_{1}\pi$, $\rho\pi$ \\
$h'_{2}$    & $2^{+-}$ &    $5-18$ &  $79$ &
$K_{1}(1400)K$, $K_{1}(1270)K$, $K^{*}_{2}(1430)K$\\ \hline 
\end{tabular}
\caption{\label{tab:exotic-hybrid-widths}
Exotic quantum number hybrid width and decay predictions.}
\end{table}

\begin{table}[h!]\centering
\begin{tabular}{|cc|cc|c|}\hline
Particle & $\mathbf{J^{PC}}$ & \multicolumn{2}{c|}{Total Width $MeV$} & 
Large Decays\\ 
         &                   & \cite{page-99} & \cite{isgur-85a} & \\ \hline
$\rho$      & $1^{--}$ & $70-121$ & $112$ & $a_{1}\pi$,$\omega\pi$, $\rho\pi$\\
$\omega$    & $1^{--}$ & $61-134$ & $60$  & $\rho\pi$, $\omega\eta$, 
$\rho(1450)\pi$\\
$\phi$      & $1^{--}$ & $95-155$& $120$ & $K_{1}(1400)K$, $K^{*}K$, 
$\phi\eta$ \\ \hline
$a_{1}$     & $1^{++}$ & $108-204$ & $269$ & $\rho(1450)\pi$, $\rho\pi$, 
$K^{*}K$\\
$h_{1}$     & $1^{++}$ & $43-130$  & $436$ & $K^{*}K$, $a_{1}\pi$    \\
$h'_{1}$    & $1^{++}$ & $119-164$ & $219$ & $K^{*}(1410)K$,$K^{*}K$ \\ \hline
$\pi$       & $0^{-+}$ & $102-224$ & $132$ & $\rho\pi$,$f_{0}(1370)\pi$ \\
$\eta$      & $0^{-+}$ & $81-210$ & $196$ & $a_{0}(1450)\pi$, $K^{*}K$ \\
$\eta'$     & $0^{-+}$ & $215-390$& $335$ & $K^{*}_{0}K$,$f_{0}(1370)\eta$,
$K^{*}K$ \\ \hline
$b_{1}$     & $1^{+-}$ & $177-338$ & $384$ & $\omega(1420)\pi$,$K^{*}K$ \\
$h_{1}$     & $1^{+-}$ & $305-529$ & $632$ & $\rho(1450)\pi$, $\rho\pi$, 
$K^{*}K$ \\
$h'_{1}$    & $1^{+-}$ & $301-373$ & $443$ & $K^{*}(1410)K$, $\phi\eta$, 
$K^{*}K$ \\
$\pi_{2}$   & $2^{-+}$ & $27-63$ & $59$ &  $\rho\pi$,$f_{2}\pi$ \\
$\eta_{2}$  & $2^{-+}$ & $27-58$ & $69$ & $a_{2}\pi$ \\
$\eta'_{2}$ & $2^{-+}$ & $38-91$  & $69$  & $K^{*}_{2}K$, $K^{*}K$ \\ \hline
\end{tabular}
\caption{\label{tab:non-exotic-hybrid-widths}
Non-exotic quantum number hybrid width and decay predictions.}
\end{table}

\begin{figure}[h!]\centering
\includegraphics[width=0.7\textwidth]{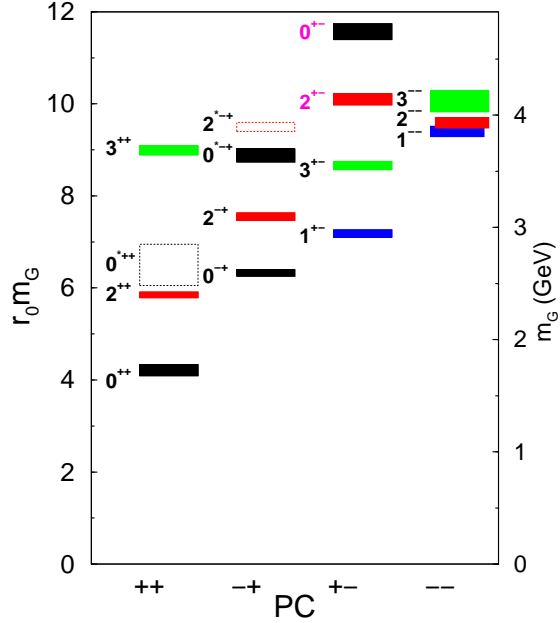}
\caption[]{\label{fig:gbspectrum}A lattice calculation of the 
glueball mass spectrum~\cite{morningstar}. The lightest three 
glueballs have quantum numbers $0^{++}$, $2^{++}$ and $0^{-+}$.
The lightest exotic glueball has quantum numbers of $2^{+-}$ with
a mass above $4$\munit .}
\end{figure}

Glueballs are nominally states of only gluons and are SU(3) singlets. The 
best predictions for the glueball spectrum comes from the lattice. A recent 
calculation using and anisotropic lattice~\cite{morningstar-99} is shown in 
Fig~\ref{fig:gbspectrum}. The lightest glueball is expected to
have $\mathbf{J^{PC}}=0^{++}$, followed by a $2^{++}$ state and
then a $0^{-+}$ state. Unfortunately, all of these quantum numbers
are  those of normal mesons. In fact the lightest glueball with 
\emph{exotic} or non--$q\bar{q}$ quantum numbers is the $2^{+-}$ near 
$4$ \munit{} and the  $0^{+-}$ state near 
$4.5$ \munit . Both well beyond the mass regime that we consider 
for light--quark mesons and deep into the charmonium region. The lightest 
glueballs will appear as an extra $f_{0}$, $f_{2}$ and $\eta$ states that
don't fit into a normal nonet. But these extra states are likely to be 
mixed with the two isoscalar states in the normal $q\bar{q}$ nonets.

If we first consider the scalar glueball, ($\mathbf{J^{PC}}=0^{++}$),
we find that the lattice prediction for the pure glueball state
is $m=(1.6\pm 0.3)$ \munit . This is extremely close to the 
nonet of scalar mesons, $a_{0}(1450)$, $f_{0}(1370)$, and $K^{*}_{0}(1430)$. 
To establish such a state as a glueball, we will first need to find
a third isoscalar state in the same mass regime. A detailed study of
productions mechanisms and decays then needs to be carried out. The naive 
predictions for the pure glueball decay to pairs of pseudoscalar mesons
is shown in Table~\ref{eq:gb-decays-1}. Under the assumption that the glueball
couples equally to all pairs of octet and singlet mesons, the following 
relationships are obtained.
\begin{equation}
\Gamma (G\rightarrow \pi\pi : K\bar{K} : \eta\eta : \eta\eta' : 
\eta' \eta' ) = 3:4:1:0:1 
\label{eq:gb-decays-1}
\end{equation}
However, in comparing glueball decays to normal mesons decays, we need to 
allow for the possibility that the glueball coupling to mesons might be
different from that of a meson coupling to  mesons. 

In looking for glueballs, there are certain production reactions which are 
considered to be \emph{glue rich}, and others that are considered to be
\emph{glue poor}. The former include $J/\psi$ decays, double-pomeron exchange
reactions and proton-antiproton annihilations. The latter include two-photon
production and photoproduction. Comparing production rates across a 
number of such reactions is crucial in establishing the gluonic nature of
an observed state.
\section{Experimental Status of Hybrids}
\begin{figure}[hp]\centering
\begin{tabular}{cc}
\includegraphics[width=0.45\textwidth]{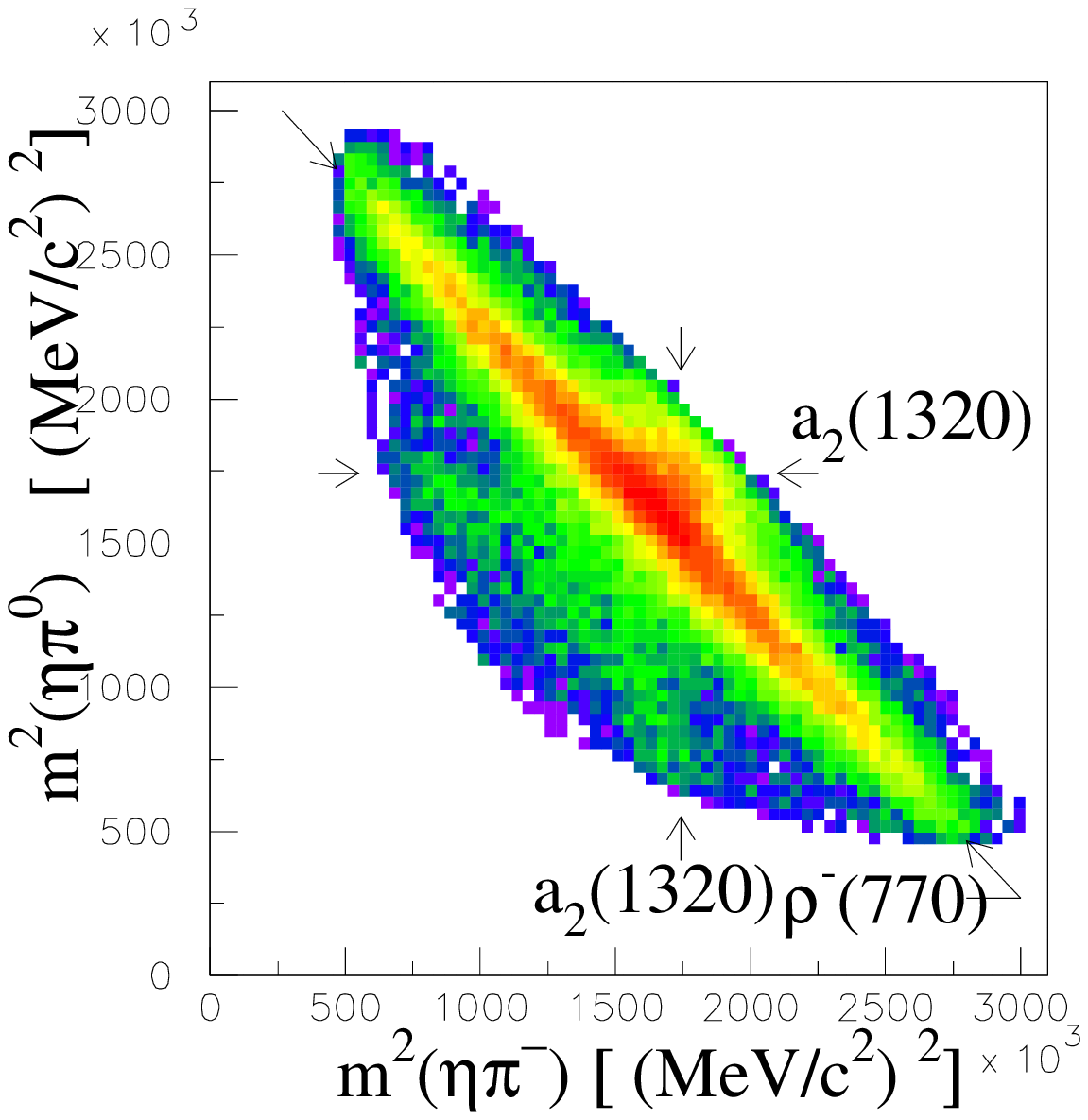} &
\includegraphics[width=0.45\textwidth]{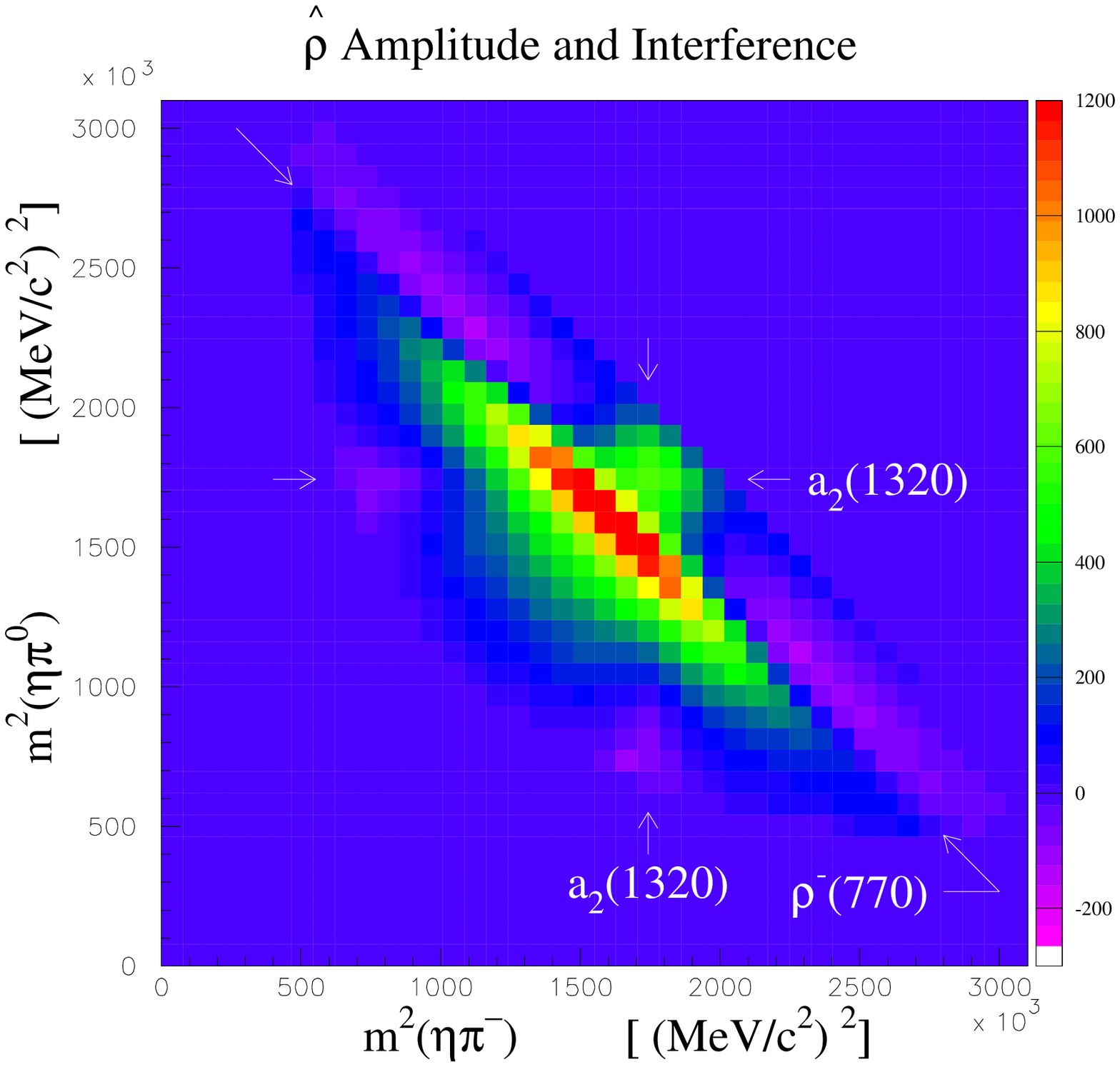} 
\end{tabular}
\caption[]{\label{fig:cbarexotic}The left hand figure shows the Dalitz
plot for $\bar{p}d\rightarrow\eta\pi^{-}\pi^{\circ}p$ with both the 
$a_{2}(1320)$ and the $\rho(770)$ indicated. The right hand figure 
shows the contribution of the $\pi_{1}(1400)$ to the Dalitz plot.}
\end{figure}

The most striking experimental prediction for hybrid mesons is the fact
that several of the nonets have non-$q\bar{q}$ quantum numbers, and
the lightest of these will be $1^{-+}$, or exotic. Over the last decade,
several credible reports of such states have been published. An isospin
$1$ object, the $\pi_{1}(1400)$ was first reported in 
$\pi^{-}p\rightarrow \eta\pi^{-} p$~\cite{e852-97}. This state was 
quickly confirmed in antiproton-neutron annihilation~\cite{Abele-98}.
Figure~\ref{fig:cbarexotic} shows the Dalitz plot from the latter 
analysis where the exotic signal is of the same strength as the 
$a_{2}(1320)$. The \textsc{pdg}~\cite{pdg-02} lists the mass as
$m=1.376\pm 0.017$ \munit{} and the width as 
$\Gamma=0.300\pm 0.040$ \munit{} with observed decays to $\pi^{-}\eta$ 
and $\pi^{\circ}\eta$.

\begin{figure}[h!]\centering
\includegraphics[width=0.75\textwidth]{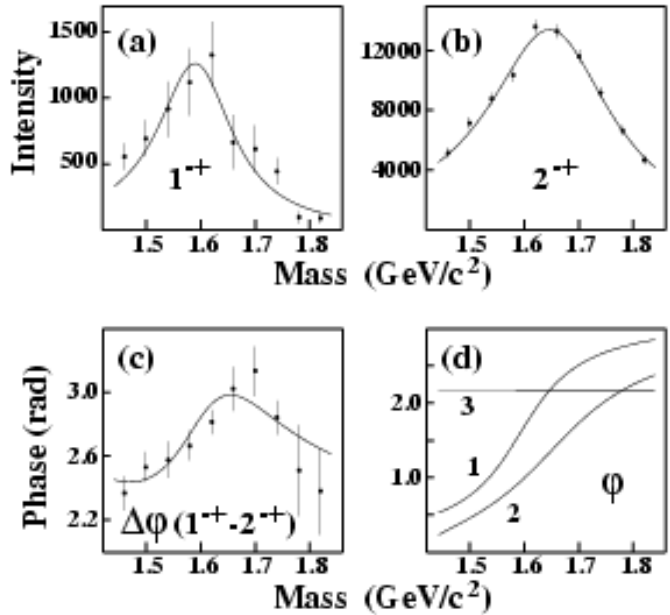}
\caption[]{\label{fig:e8523p}Data from the E852 experiment showing and
exotic $1^{-+}$ signal in the $\rho\pi$ subsystem of 
$\pi^{-}p\rightarrow \pi^{+}\pi^{-}\pi^{-}p$. (a) shows the intensity
of the $1^{-+}$ signal which interferes with (b) the $2^{-+}$ $\pi_{2}(1670)$.
(c) shows the phase difference between the two waves, and (d) has the 
individual phases, with $1$ corresponding to the $1^{-+}$, $2$ to the 
$2^{-+}$ and $3$ to a background term.}
\end{figure}

A second such state, the $\pi_{1}(1600)$, was first observed in
$\pi^{-}p\rightarrow \pi^{+}\pi^{-}\pi^{-} p$~\cite{e852-98}.
The signal for the $\pi(1600)$ is shown in Figure~\ref{fig:e8523p}.
A latter observation reported the $\pi_{1}(1600)\rightarrow\eta ' \pi$
\cite{e852-01} and various reports have been made at conferences about 
other observed decay modes. The VES experiment~\cite{ves} reports the
ratios of: 
$b_{1}\pi : \eta' \pi : \rho\pi\,\,=\,\, 1:1.0\pm 0.3 : 1.6\pm 0.4$. 
The \textsc{pdg}~\cite{pdg-02} lists the mass as 
$m=1.596^{+0.025}_{-0.014}$ \munit{} and the width as 
$\Gamma=0.312^{+0.064}_{-0.024}$ \munit . Recently there has been a 
report of the $f_{1}(1285)\pi$ decay mode of the $\pi(1600)$ as well third 
$\pi_{1}$ state in the $1.9$ \munit{} region, also decaying to 
$f_{1}\pi$~\cite{weygand}.

The precise interpretation of these states is still open. The $\pi_{1}(1400)$
is significantly lighter than theoretical expectations, and its only
observed decay mode, $\eta\pi$ is not expected for a hybrid. Recent
work suggests that this state may actually be non-resonant scattering
similar to the S-wave $\pi\pi$ scattering at low energy~\cite{dzierba-03}.
The same explanation in the $\pi\eta'$ system can also be invoked
to explain a large part of the $\eta'\pi$ signal for the 
$\pi_{1}(1600)$~\cite{szczepaniak-03}. 
The $\pi_{1}(1600)$ as seen in $\rho\pi$ is
still somewhat lower than theoretical expectations in mass, but could
well be a hybrid meson. The open question now is what are its decay 
modes, and can we find any of its partner states, $\eta_{1}$ and 
$\eta_{1}'$? 

There is also a more general issue of what is causing the over population
of $\pi_{1}$ states? There is one $1^{-+}$ hybrid nonet, meaning that there
should only be one $\pi_{1}$ state. While it is possible that the 
$\pi_{1}(1400)$ is just final state interactions, if there are really
two states beyond this, ($\pi_{1}(1600)$ and $\pi_{1}(1900)$), it will 
be necessary to rethink what is happening. 

Hybrids with non-exotic quantum numbers are more difficult to 
discern as they look like normal $q\bar{q}$ mesons. 
Table~\ref{tab:non-exotic-hybrid-widths} shows the widths and decays
expected for these states from model calculations. If one assumes that
the $\pi_{1}(1600)$ sets the mass scale for the hybrids, then we are looking
in the $1.6$ to $2.2$ \munit{} mass range. 

In the $J^{PC}=0^{-+}$ system, we expect radial excitations of the 
pseudoscalar mesons as well as a glueball state. Three states of 
interest appear in this sector, the $\pi(1800)$, ($m=1.8$, $\Gamma=0.21$)
has been observed with decays into $f_{0}(980)\pi$, $f_{0}(1370)\pi$,
$\rho\pi$, $\eta\eta\pi$, $a_{0}(980)\pi$ and $f_{0}(1500)\pi$. and there
has been speculation that due to its coupling to scalars, it may have
a large hybrid component. The  $\eta(1760)$, which decays into $4\pi$, has 
only been observed in $J/\psi$ decays. This is a likely partner for the
$\pi(1800)$. Finally, the $\eta(2225)$, ($m=2.2$,$\Gamma=0.15$) has been
observed in $J/\psi$ with decays into $\phi\phi$. This state is too high
in mass to be the simple partner of the other two, but is consistent with
what is expected of a glueball but needs confirmation.

In the $J^{PC}=1^{--}$ system, we expect to see the radial excitations of
the vector mesons as well as the $^{3}D_{1}$ nonet.  The mass scale for
the D-wave mesons are set by the $^{3}D_{3}$ nonet, $\rho_{3}(1690)$,
$\omega_{3}(1670)$ and the $\phi_{3}(1850)$. There are a rather large
number of known states in this region. The $\rho(1450)$, $\rho(1700)$,
$\rho(1900)$, $\rho(2150)$, $\omega(1420)$, $\omega(1650)$, and 
$\phi(1680)$.  This sector is probably completely mixed, so disentangling 
it is going to require a clear understanding of other sectors. 

The $J^{PC}=1^{+-}$ hybrid nonet is near the radial excitations
of the $b_{1}$s nonet. One known state exists, the $h_{1}(1595)$, 
($m=1.6$, $\Gamma=0.38$), but little is known it. It is
probably consistent with being a  radial excitation of the $h_{1}(1170)$.

The $J^{PC}=1^{++}$ nonet has the same quantum numbers as the radial 
excitations of the $a_{1}$s nonet. One known state exists, the 
$a_{1}(1640)$ which has been observed in a $3\pi$ decay. What little is 
known is consistent with this being a radial excitation of the $a_{1}(1260)$.

The $J^{PC}=2^{-+}$ nonet can overlap with the D-wave nonet, $^{1}D_{2}$.
There are a rather large number of candidates here. The $\pi_{2}(1670)$
and the $\eta_{2}(1645)$ are reasonably consistent with the D-wave mesons.
There is a second $\eta_{2}$, the $\eta_{2}(1870)$ that mass-wise is 
consistent with being the $\eta'$ of this nonet. However, its decay 
modes are consistent with it being composed of mostly non-strange light
quarks. In fact, the $a_{2}\pi$ decay appears to be the largest mode
for both the $\eta_{2}(1645)$ and the $\eta_{2}(1870)$. This sector
has the strongest evidence for a hybrid state. 

In any case, establishing the non-exotic hybrid nonets will almost
certainly require more exotic states. These will allow us to both
set the mass scale, and understand the actual decay patterns of the
states.

\section{Experimental Status of Glueballs}

\begin{figure}[h!]\centering
\begin{tabular}{cc}
\includegraphics[width=0.45\textwidth]{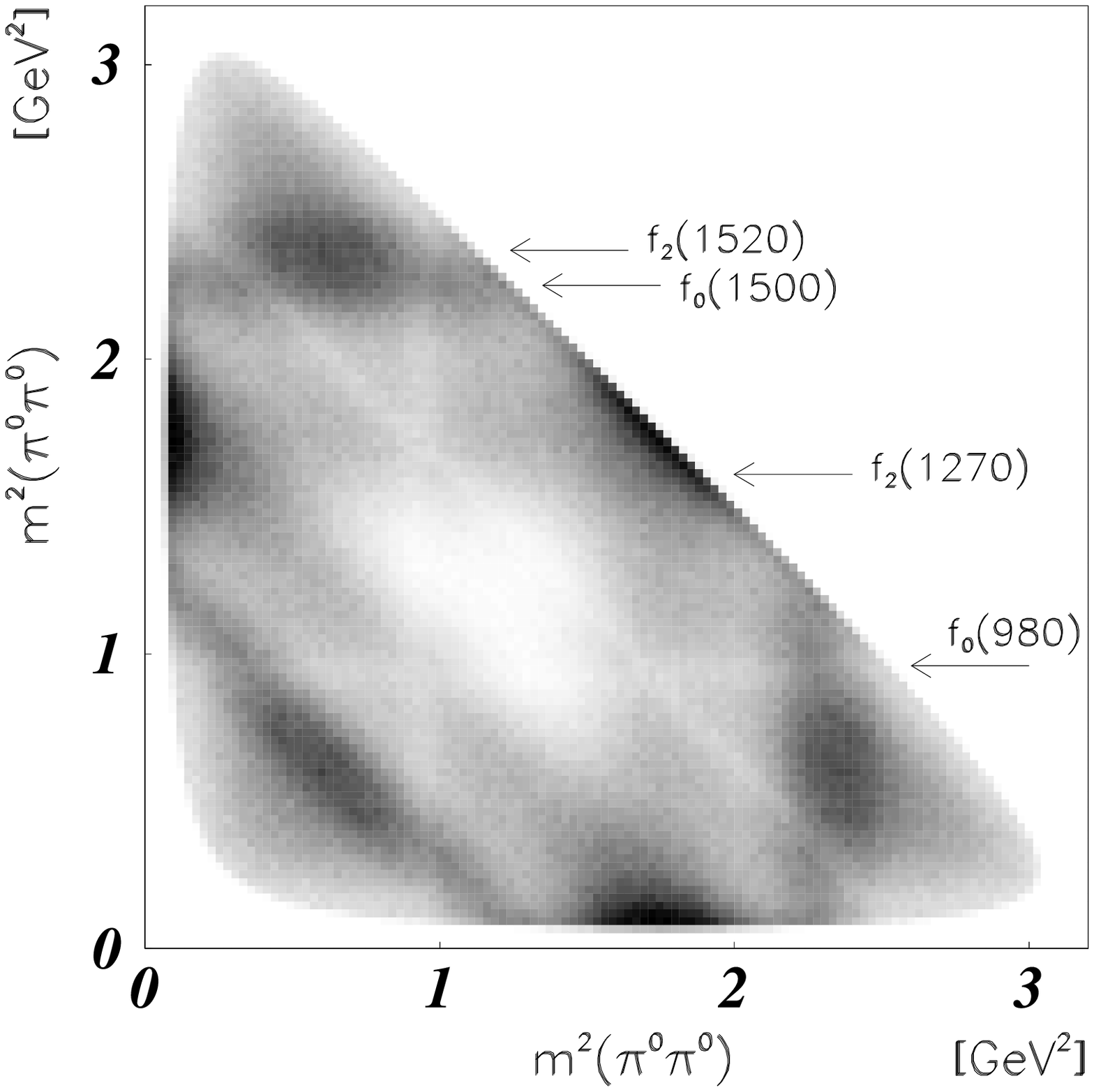}
\includegraphics[width=0.45\textwidth]{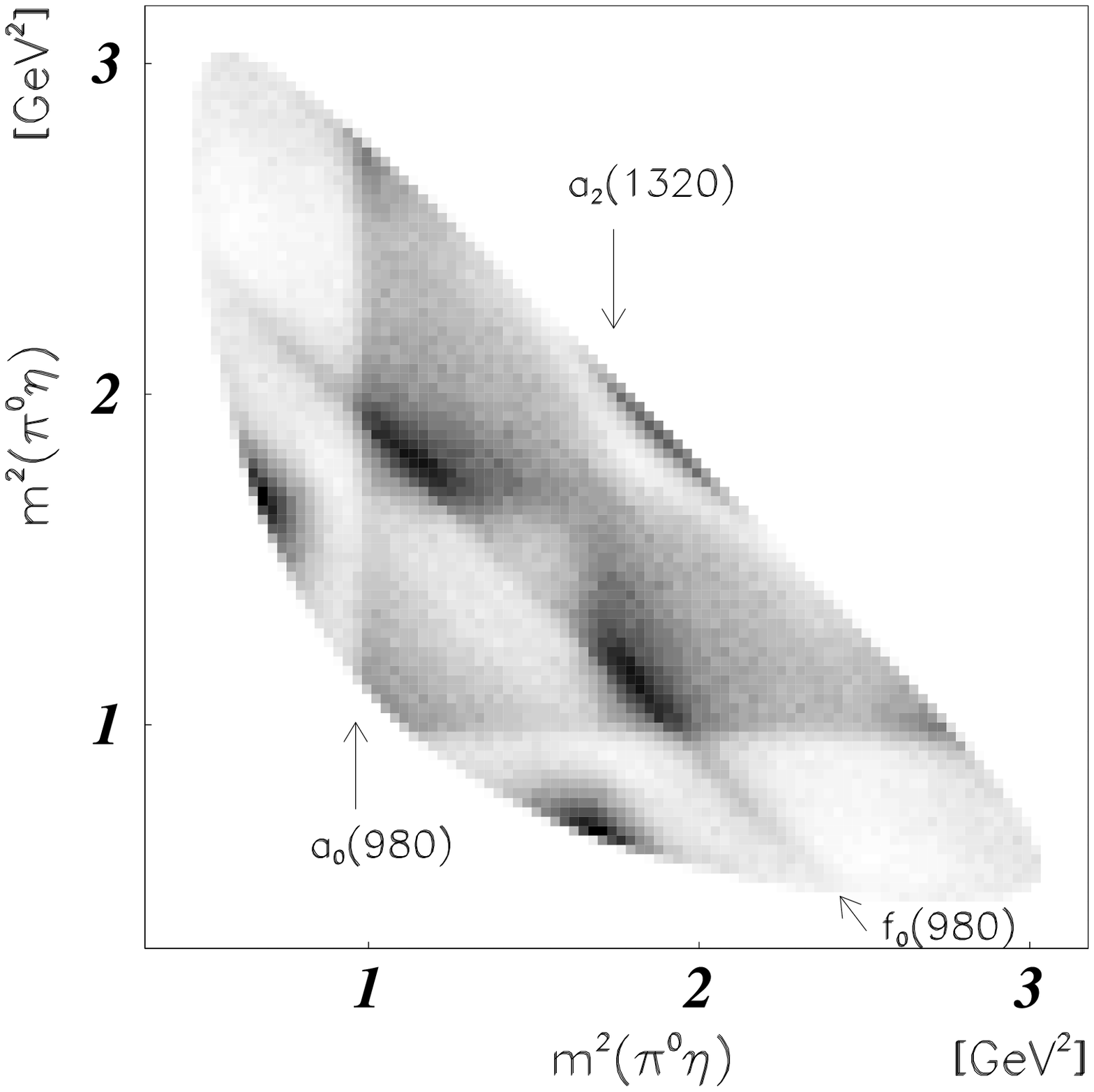}
\end{tabular}
\caption[]{\label{fig:dal-1}Dalitz plots for $\bar{p}p$ annihilation
at rest into $3\pi^{\circ}$ (left) and $2\pi^{\circ}\eta$ (right). The
resonances as observed in the partial wave analysis are indicated on the
figure.}
\end{figure}

In Fig.~\ref{fig:dal-1} are shown the Dalitz plots for $p\bar{p}$
annihilation at rest into $\pi^{\circ}\pi^{\circ}\pi^{\circ}$ and
$\pi^{\circ}\eta\eta$. While the analysis of these channels involve
many intermediate resonances~\cite{amsler-95a,amsler-95c}, there
is one new state which stands out in both, the $f_{0}(1500)$.
This state has a mass of $1.505$ \munit{} and a width of
$0.110$ \munit . In later analysis~\cite{amsler-95b,amsler-94a},
the $f_{0}(1500)$ has also been observed in the $\eta\eta'$ and
$K_{L}K_{L}$ final states.. Examining all these data, it is possible 
to extract many different annihilation--decay rates for the $f_{0}(1500)$ 
as given in Table~\ref{tab:f01500}. We can convert the numbers in this 
table into relative decay amplitudes squared, $\gamma^{2}$, which if 
normalized to the $\eta\eta$ mode are yield the rates given in 
Table~\ref{tab:ratios}. These rates are not consistent with either a
pure mesons or a pure glueball and yielded the first evidence that 
something interesting is happening in the scalar sector.

\begin{table}[h!]
\begin{tabular}{|l|c|c|} \hline
\textsl{Decay} & \textsl{Rate} & $q\,\mathrm{\left[GeV/c\right]}$ \\ \hline
$f_{0}(1500)\rightarrow\pi\pi$            & $0.290\pm 0.075$ & $0.740$ \\ 
$f_{0}(1500)\rightarrow K\bar{K}$         & $0.035\pm 0.003$ & $0.567$  \\ 
$f_{0}(1500)\rightarrow\eta\eta$          & $0.046\pm 0.013$ & $0.516$ \\ 
$f_{0}(1500)\rightarrow\eta\eta^{\prime}$ & $0.012\pm 0.003$ & $0.0889$  \\ 
$f_{0}(1500)\rightarrow 4\pi$             & $0.617\pm 0.096$ & $0.5$\\ \hline
\end{tabular} 
\caption{\label{tab:f01500}
Measured branching fractions for 
$\bar{p}p\rightarrow f_{0}(1500)\pi^{\circ}$.}
\end{table}
Another glue-rich channel is that of central production, and a great
deal of analysis has been done recently by the WA102 collaboration
at CERN. They  have looked at central production of 
$\pi^{+}\pi^{-}$~\cite{barberis-95,barberis-99b}, 
$\pi^{\circ}\pi^{\circ}$~\cite{barberis-99c},
$K\bar{K}$~\cite{barberis-99a}, $\eta\eta$~\cite{barberis-00a} and 
$\pi^{+}\pi^{-}\pi^{+}\pi^{-}$~\cite{barberis-95,barberis-97b,barberis-00b} in
$450\,\mathrm{GeV/c}$ $pp$ collisions. In all of these analysis,
they observe two scalar states, the $f_{0}(1500)$ and the $f_{0}(1710)$.
In addition, in the $4\pi$ data, they observe the $f_{0}(1370)$.
They also find that by kinematically selecting on 
their data, they were able to enhance the scalar signals.
\begin{table}[h!]\centering
\begin{tabular}{|c|c|c|c|c|}\hline
State & Mass \munit{} & Width \munit{}  
& Decay & Reference \\ \hline
$f_{0}(980)$ & $0.985\pm 0.010$ & $0.065\pm0.020$ & $K^{+}K^{-}$ & \cite{barberis-99a} \\
$f_{0}(980)$ & $0.982\pm 0.003$ & $0.080\pm0.010$ & $\pi^{+}\pi^{-}$ & \cite{barberis-99b} \\
$f_{0}(1370)$ & $1.290\pm 0.015$ & $0.290\pm 0.030$ & 
$\pi^{+}\pi^{-}\pi^{+}\pi^{-}$ & \cite{barberis-97b} \\
$f_{0}(1370)$ & $1.308\pm 0.010$ & $0.222\pm 0.020$ & 
$\pi^{+}\pi^{-}$ & \cite{barberis-99b} \\
$f_{0}(1500)$ & $1.502\pm 0.010$ & $0.131\pm 0.015$ & 
$\pi^{+}\pi^{-}$ & \cite{barberis-99b} \\
$f_{0}(1500)$ & $1.497\pm 0.010$ & $0.104\pm 0.025$ & 
$K^{+}K^{-}$ & \cite{barberis-99a} \\
$f_{0}(1500)$ & $1.510\pm 0.020$ & $0.120\pm 0.035$ & 
$\pi^{+}\pi^{-}\pi^{+}\pi^{-}$ & \cite{barberis-97b} \\
$f_{0}(1710)$ & $1.700\pm 0.015$ & $0.100\pm 0.025$ & 
$K^{+}K^{-}$ & \cite{barberis-99a} \\ 
$f_{0}(2000)$ & $2.020\pm 0.035$ & $0.410\pm 0.050$ & 
$\pi^{+}\pi^{-}\pi^{+}\pi^{-}$ & \cite{barberis-97b} \\ \hline
\end{tabular}
\caption{\label{tab-wa102}
Observed scalar mesons in various final states in WA102.}
\end{table}
\begin{table}[h!]\centering
\begin{tabular}{|l|c|c|c|c||c|}\hline
{\sf Decay Rate} & $\pi\pi$ & $\bar{K}K$ & $\eta\eta$ & $\eta\eta^{\prime}$ & 
$4\pi$ \\ \hline
$f_{0}(1500)$    & $5.13\pm 1.95$ & $0.708\pm 0.209$ & $1.00$ & $1.64\pm 0.62$ &
$13.7\pm 4.4$ \\ \hline
{\sf Meson}      & $14$ & $1.4$ & $1$ & $2.4$ &  \\ \hline
{\sf Glueball}   & $3$ & $4$ & $1$ & $0$ & {\sf Large} \\ \hline
\end{tabular} 
\caption{\label{tab:ratios}
Relative decay amplitudes squared, $\gamma^{2}$
normalized to the $\eta\eta$ rate for the $f_{0}(1500)$. These are compared
to the SU(3) prediction for an $s\bar{s}$ meson with mixing angle of
$150^{\circ}$, as well as for a pure glueball.}
\end{table}

\begin{figure}
\begin{tabular}{ccc}
\includegraphics[width=0.30\textwidth]{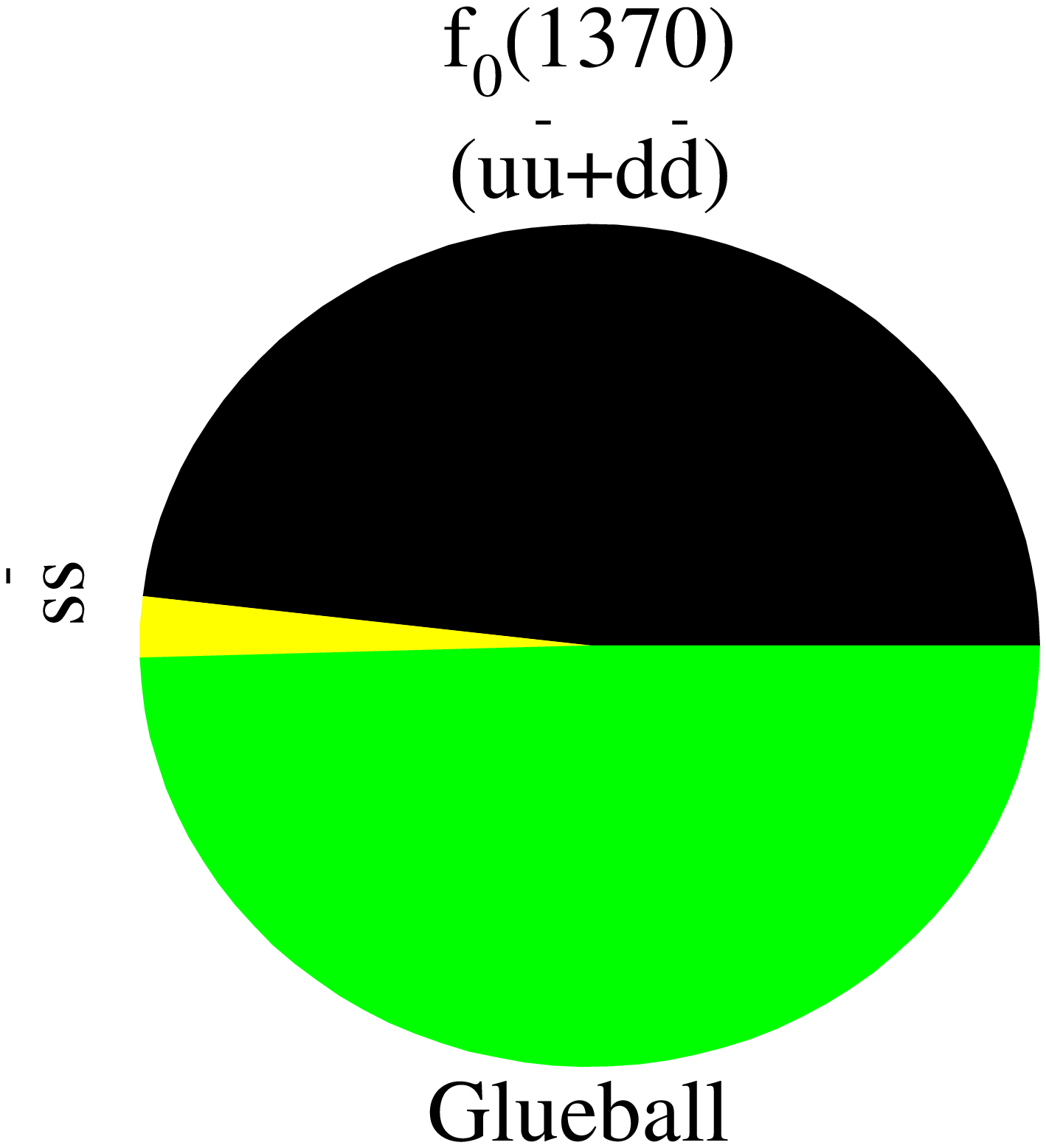} &
\includegraphics[width=0.30\textwidth]{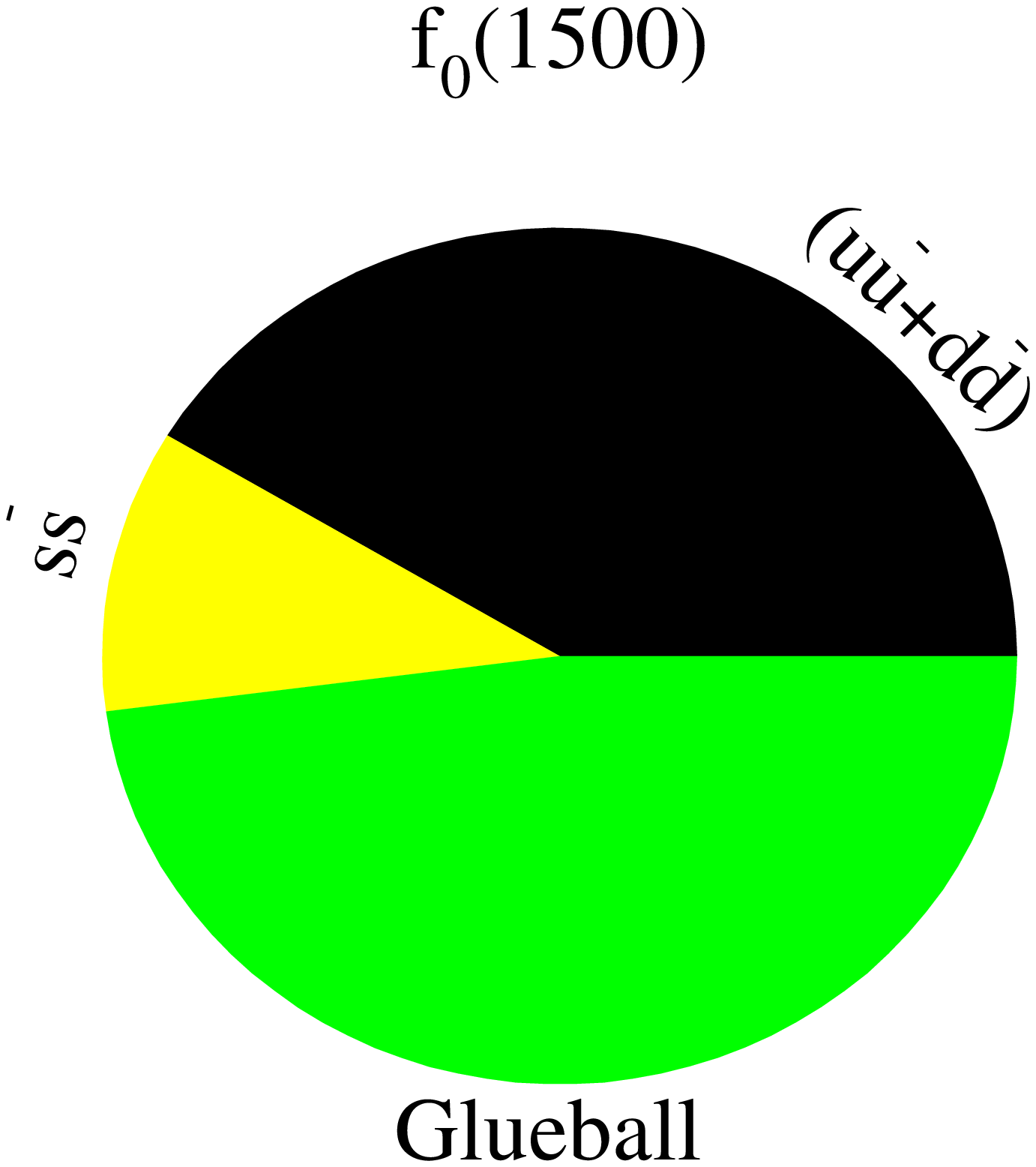} &
\includegraphics[width=0.30\textwidth]{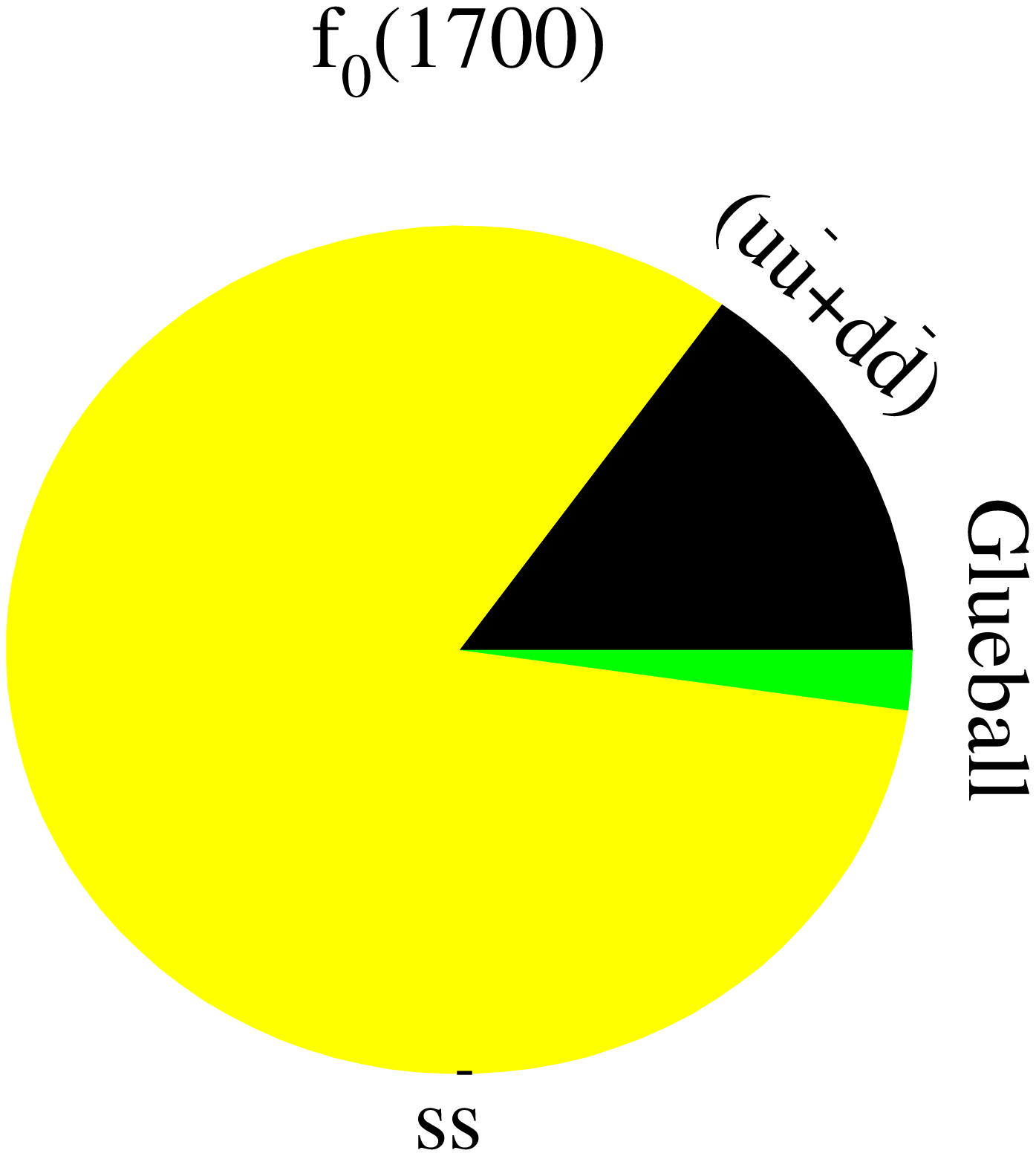}
\end{tabular}
\caption[]{\label{fig:mixing}The meson and glueball content of the 
$f_{0}(1370)$, $f_{0}(1500)$ and $f_{0}(1710)$ states.}
\end{figure}

A recent analysis of all available information on scalar 
decays~\cite{close-01} comes up with a plausible mixing scenario that 
explains the $f_{0}(1370)$, $f_{0}(1500)$ and $f_{0}(1710)$. It is based 
on a model that allows for three different decay couplings, and possible
flavor violations in glueball decays. This analysis finds bare 
masses for the scalar glueball of $m_{G}=1.443\pm 0.024$\munit{},
that of the $u\bar{u}/d\bar{d}$ meson of $m_{N}=1.377\pm 0.020$\munit{}
and the $s\bar{s}$ meson at $m_{S}=1.674\pm 0.010$\munit . The mixing scheme
for each of the three physical states is given below and is graphically
shown in Figure~\ref{fig:mixing}. 
\begin{eqnarray*} 
\begin{array}{c}
\\ 
f_{0}(1710) \\
f_{0}(1500) \\
f_{0}(1370) 
\end{array} & \,\,\,\, & \begin{array}{ccc} 
 f_{i1}^{(G)}  & f_{i2}^{(S)} & f_{i3}^{(N)}  \\
  (0.39\pm 0.03) & (0.91\pm 0.02) & (0.15\pm  0.02) \\
 (-0.65\pm 0.04) & (0.33\pm 0.04) & (-0.70\pm 0.07) \\
 (-0.69\pm 0.07) & (0.15\pm 0.01) & (0.70\pm 0.07) 
\end{array}
\end{eqnarray*}
While the precise mixing angles may
not be fully pinned down, there are some interesting trends. The central
mass state, $f_{0}(1500)$ has $s\bar{s}$ and $u\bar{u}/d\bar{d}$ out of 
phase, while the upper and lower states have them in phase. The glueball 
component is in-phase with the SU(3) singlet piece for the $f_{0}(1710)$, 
and out of phase for the $f_{0}(1370)$. Lastly, the glueball is distributed 
over all three states. The upper state is mostly $s\bar{s}$, while the 
lower two contain mostly $u\bar{u}$ and $d\bar{d}$.

While there appears to be solid evidence for a scalar glueball mixed
into the scalar nonet, the evidence for the $2^{++}$ and $0^{-+}$ glueballs
is significantly weaker. The tensor state is supposed to be next lightest
with a predicted mass of about $2.2$\munit . A possible candidate, the
$f_{2}(1950)$ has been observed in central production in a similar 
fashion to the scalar glueball candidates~\cite{barberis-00b}, although
there is little other information about this state. A second state,
the $\xi(2220)$ was originally reported in $J/\psi$ radiative decays.
This very narrow object ($\approx 0.02$\munit ) was originally seen by 
one experiment~\cite{baltrusaitis-86}, but not confirmed by several others.
No spin parity analysis is available, but $2^{++}$ is favored.
It was later seen in the BES experiment~\cite{bai-98} with approximately
equal strength in $\pi\pi$, $\bar{K}K$ and $\bar{p}p$ annihilations. Later
careful searches in $\bar{p}p$ annihilation~\cite{amsler-01} showed no
evidence of this state. These data lead to a lower limit of 
$J/\psi\rightarrow\gamma\xi$ of 
$\approx 2.3\times 10^{-3}$~\cite{godfrey-99}. If this state really exists, 
it has one of the largest radiative branching fractions in $J/\psi$ decays. 
The existence of this state is still an open question. Evidence for the 
$0^{-+}$ glueball is even weaker, and full exploration of the glueball 
sector awaits new data.

\section{Summary}
Clear experimental evidence exists for both mesons with non-$q\bar{q}$
quantum numbers and for a scalar glueball which is mixed with the nearby
scalar mesons. Unfortunately, the exact nature of the exotic states remain
unclear. Their observed mass and decay modes do not completely agree
with theoretical expectations, which may well indicate that there are
problems with the theory. In order to resolve these issues, it will be
necessary to observe and measure both the partners of the existing states
as well as states with other exotic quantum numbers. This will be studied
in the light quark sector with the GlueX Experiment at Jefferson 
Lab~\cite{gluex}
and in the charmonium sector with the PANDA experiment at GSI~\cite{peters}.

While there is a good explanation of the scalar glueball, a full 
understanding of the glueballs will require the discovery and 
study of at least one additional state. There is a very good chance that
this will be accomplished with the CLEO-c experiment in the near 
future~\cite{alexander}.

\end{document}